\documentclass[aps,showpacs]{revtex4}
\usepackage{fullpage,amsfonts,amssymb,epsfig,graphicx}

\begin{document}

\title{Kinematical classification of two-pion production on the nucleon}
\author{N.E. Ligterink}
\affiliation{
Department of Physics and Astronomy, University of
Pittsburgh, \\
3941 O'Hara Street, Pittsburgh, PA 15260, U.S.A.}
\begin{abstract}
We give a full kinematical classification of all the tree-level
two-pion photoproduction processes on the nucleon, which consists
of seventeen diagrams. It suggests a method
of analysis of two-pion data with little model bias.
\end{abstract}

\maketitle

\section{Introduction}

Theoretically, photo- and electroproduction of two pions on
the nucleon is notoriously difficult due to the complexity
and numerous processes involved. A serious attempt would
require close to a hunderd distinct diagrams. Model 
assumptions often appear in the data analysis by the restricted
choice of processes. Therefore it is necessary to make a 
systematic study, and catagorizes these processes. This paper 
has the limited goal of classifying all the processes by
their kinematical properties and showing their characteristics,
similar to the $s$-, $t$-, and $u$-channel distinction of
2-to-2 scattering.

The two-pion production on the nucleon plays a pivotal role
in the analysis of baryon resonances in the intermediate
energies as it is the dominant final state. However, the 
analysis is often restricted. Most of the time the analysis
is performed on integrated quantities, where only a subset
or a projection of the five-dimensional data is used. Experiments
with almost $4 \pi$ acceptance would still have problems to fill
the last gaps at forward angles, where certain $t$-channel 
processes are important.
Furthermore, the analysis is often of the type of an isobar model,
which focuses on $s$-channel processes.

In the two-pion data the contact, or Kroll-Ruderman, term plays a
dominant part. It is the contact photo-production of a pion 
on the nucleon, and has the structure given by the $\pi NN$
coupling where the derivative in the interaction
$i \partial^\mu $ is replaced by the minimal photon coupling 
$e A^\mu$. The nucleon can be left in the excited state.
For example, Murphy and Laget \cite{Murphy:1996ms} included 
$\Delta$ and $N^\ast$ excitations
for the nucleon, which decayed under the emission of a
second pion.

The width of a resonance can only be reproduced by
an infinite summation of a perturbative series. From 
this and the strength of the
interaction it can be argued that a perturbative approach has a
limited applicability, especially once the energy is larger, such
as for the second resonance region around 1.6 GeV.
However, G\'omez Tejedor and Oset \cite{GomezTejedor:1995pe} 
did pursue chiral
perturbation theory to analyse different isospin
channels up to this energy. They were able to trace
certain channels back to particular baryon resonances.
A similar, but more limited analysis was performed by
Ochi, Hirata, and Takaki. \cite{Ochi:1997ev,Hirata:gr}
The Valencia group also restricted 
itself to particular subprocesses and the structure of
the $N(1520)$ resonance \cite{GomezTejedor:1995kj}
and the $\Delta(1700)$ resonance\cite{Nacher:2000eq}
as seen in the two-pion production.

Although polarization is important to distinguish 
different angular momentum states, and in, for example,
the Gerasimov-Drell-Hearn sum rule \cite{Nacher:2001yr},
the polarized large acceptance data is not common.
Hence at the moment it seems more important to focus
on kinematical separation of the different production
processes, which we do in this short paper.

\begin{figure}
\centerline{\includegraphics[width=14cm]{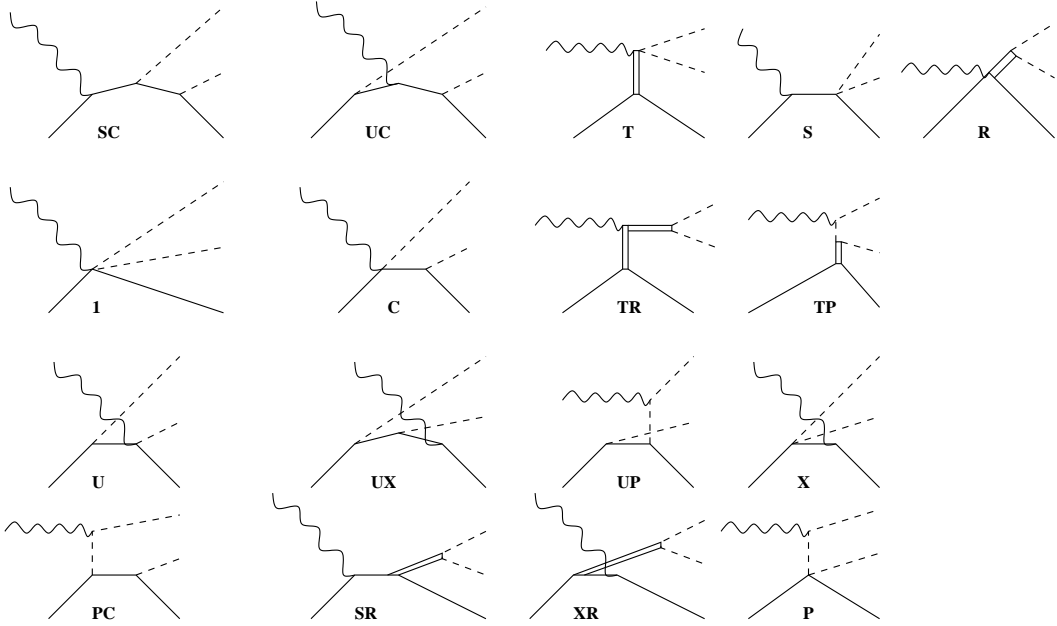}}
\caption{The seventeen tree-level production processes
labeled by their intermediate states. The solid lines
are all possible baryons, the double lines all possible
mesons, and the dashed lines are pions.} \label{fig1}
\end{figure}

\section{Theory}

If one lumps all the baryon and baryonic excitations together,
and do the same with all the mesonic states one is still left with
seventeen possible production tree diagrams, which are shown in
Fig.~\ref{fig1}. We ignored the Vector Meson Dominance diagrams
where the photon oscillates to a rho-meson, which we consider
a vertex function, or form factor. These diagrams, apart from the
possible vertex
functions we will discuss later, depend only on a restricted
set of invariant momenta, which are indicated by the letters in 
the figure. If we denote the incoming nucleon four-momentum by
$p_{ni}$, the outgoing nucleon momentum by $p_{nf}$, and
the two pion momenta by $p_{\pi 1}$ and $p_{\pi 2}$, we have the
following set of functions of a single momentum squared:
\begin{eqnarray}
s{\rm -channel\ production} & : & S(p_{nf}+p_{\pi 1}+p_{\pi 2}) \ \ ,\\ 
t{\rm -channel\ production} & : & T(p_{nf}-p_{ni}) \ \ ,\\ 
u{\rm -channel\ production} & : & U_\alpha(p_{ni}-p_{\pi \alpha}) \ \ ,\\ 
{\rm cascade;\ decay\ product} & : & C_\alpha(p_{nf}+p_{\pi \alpha}) \ \
,\\ 
\pi-\pi{\rm  -channel} & : & R(p_{\pi 1}+p_{\pi 2}) \ \ ,\\ 
\pi{\rm -photon\ interaction} & : & P_\alpha(p_{nf} +p_{\pi \alpha} -
p_{ni}) \ \ ,\\  
{\rm late\ photon\ absorption} & : & X(p_{ni} - p_{\pi 1}- p_{\pi 2}) \
\ .
\end{eqnarray}
If the pions are identical one should sum $p_{\pi 1} \leftrightarrow
p_{\pi 2}$, otherwise there are two functions $\alpha=1,2$ for
$U_\alpha$, $C_\alpha$, and $P_\alpha$.
Excluding final-state interactions, the amplitude is the 
particular linear combination of the functions of 
these variables, shown in Fig.~\ref{fig1}.
Furthermore, the functions $S, U, X,$ and $C$ can only contain
the baryonic resonances, while the functions $P, R,$ and $T$ only the
mesonic resonances. As they stand, the kinematical diagrams, which
contain only the denominator of the propagators, indicated by a single 
letter have the dimension of $[{\rm energy}]^{-2}$ and the double
letter diagrams $[{\rm energy}]^{-4}$, together with vertex functions and
dimensionful coupling constants they should yield a
dimensionless invariant amplitude.

Of course, a careful spin, isospin, and partial wave analysis will 
exclude particular combinations for spin-isospin quantum
numbers of the final state. For example, the neutral pion
will not generate photo-pion interactions, hence, that amplitude
will not depend on $P$. However, at the moment we are more interested
to determine the kinematical signatures of each of the processes. It has
been the custom to ignore the $t$- and $u$- processes, however,
at the typical scattering energies nowadays where pions are
fully relativistic, there is no a priori reason to do so. 

It is difficult to represent five-dimensional differential cross sections. 
For two-pion
data one often plots the Dalitz plot, which is the density of events
for a given invariant mass $m^2_{n1}= (p_{nf}+p_{\pi 1})^2$ and
$m^2_{n2}= (p_{nf}+p_{\pi 2})^2$. These two variables determine the
length of each of the 3-momentum $|{\bf p}_{nf}|$, $|{\bf p}_{\pi 1}|$,
and $|{\bf p}_{\pi 2}|$, in a given frame. However, 
their orientation with respect
to the initial scattering direction ${\bf p}_\gamma$ is not fixed.
For unpolarized scattering there are two angles, which give the
rigid-body rotation of the triangle ${\bf p}_{nf}-{\bf p}_{\pi 1}-
{\bf p}_{\pi 2}$ in the center-of-mass frame with respect to the vector
${\bf p}_\gamma$, while for a polarized
photon there is an additional angle with respect to the polarization
direction ${\bf e}_\gamma$. For unpolarized production processes in
the $s$-channel ($S$), the amplitude is constant for these angles.
For all the others there is an angular dependence, which might be weak.

The propagators are taken to be polarization averaged. If all the 
spin projections of a state are summed, it yields a projection
on the relative momenta $k_1-k_2$ of the decay products orthogonal to the 
momentum $p$ of the decaying particle. Hence
\begin{equation}
\Gamma^{(l)}(p,k_1,k_2) =
\langle p^0, {\bf p}| k^0_1,{\bf k}_1; k^0_2,{\bf k}_2 \rangle  =
\left(((k_2-k_1)\cdot p)^2 -p^2 (k_2-k_1)^2\right)^{l/2}\ \ .
\end{equation}

\begin{figure}
\centerline{
\includegraphics[width=8cm]{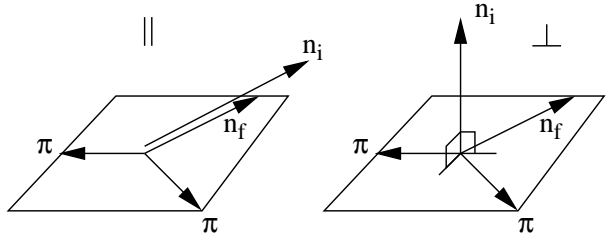}}
\caption{The two kinematical cases studied here, given in the
center-of-mass frame. Parallel kinematics ($\parallel$)
means the incoming and outgoing nucleon have their
momentum in the same direction. Perpendicular kinematics
means that the incoming nucleon momentum is perpendicular to
both the outgoing nucleon and pion momenta.}\label{fig2}
\end{figure}

The general effect of such a vertex factor is a suppression at 
threshold of the decay product. It is not difficult to insert factors 
like these in the invariant amplitudes, however, since we are interested
in a general classification only, we use unit vertex
functions everywhere. For each channel we insert a single resonance:
\begin{eqnarray}
S(p^2) & = & \frac{1}{p^2 - M_s^2 + i \Gamma_s} \ \ , \\
C(p^2) = U(p^2) = X(p^2) & = & \frac{1}{p^2 - M_c^2 + i \Gamma_c} \ \ , \\
T(p^2) = R(p^2) & = & \frac{1}{p^2 - m_r^2 + i \Gamma_r} \ \ , \\
P(p^2) & = & \frac{1}{p^2 - m_\pi^2 + i \epsilon} \ \ , 
\end{eqnarray}
where we have taken the general idea of a second, or higher
resonance in the $s$-channel, a lower resonance
in the $c$, $x$, and $u$-channel, and a meson
resonance in the $t$ and $r$ channel. The
pion is taken as an elementary particle in the 
photon interaction. The invariant amplitude
corresponds to the complex multiplication
of the corresponding functions.
To investigate the angular dependence, we study two cases:
the incoming nucleon parallel to the outgoing nucleon ($\parallel$),
and the incoming nucleon perpendicular to the outgoing
momenta in the center-of-mass frame ($\perp$). (See Fig.~\ref{fig2}.)
This does not exhaust the angular dependence, but will
give a good indication for each of the processes.

\section{Results}

\begin{figure}
\centerline{
\includegraphics[width=8cm]{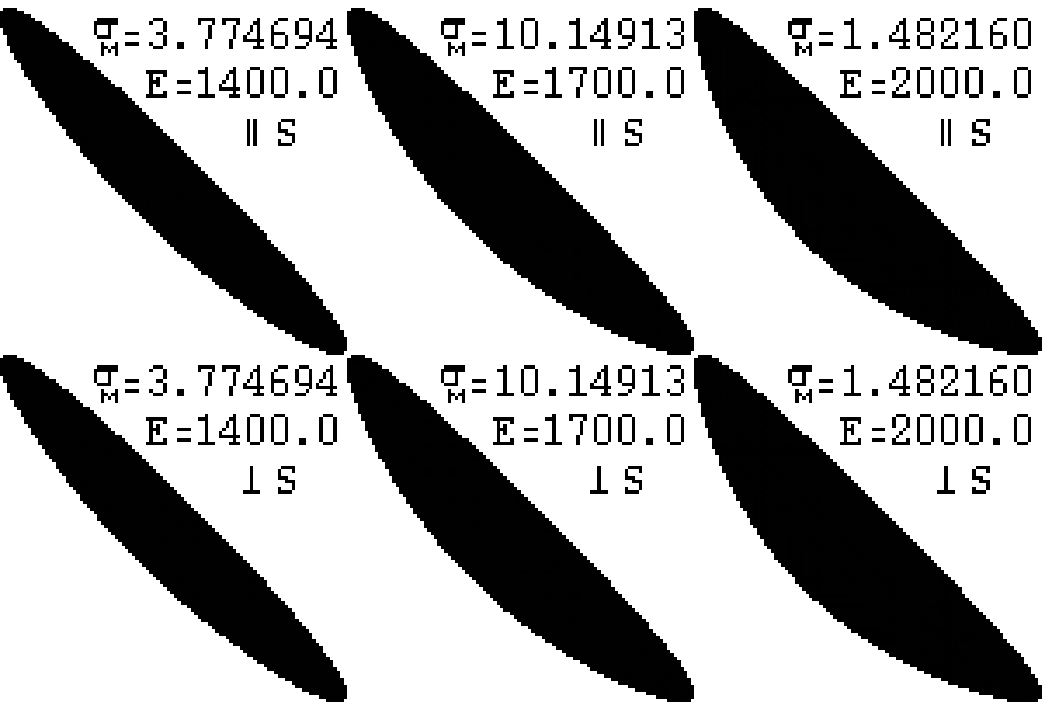}\includegraphics[width=8cm]{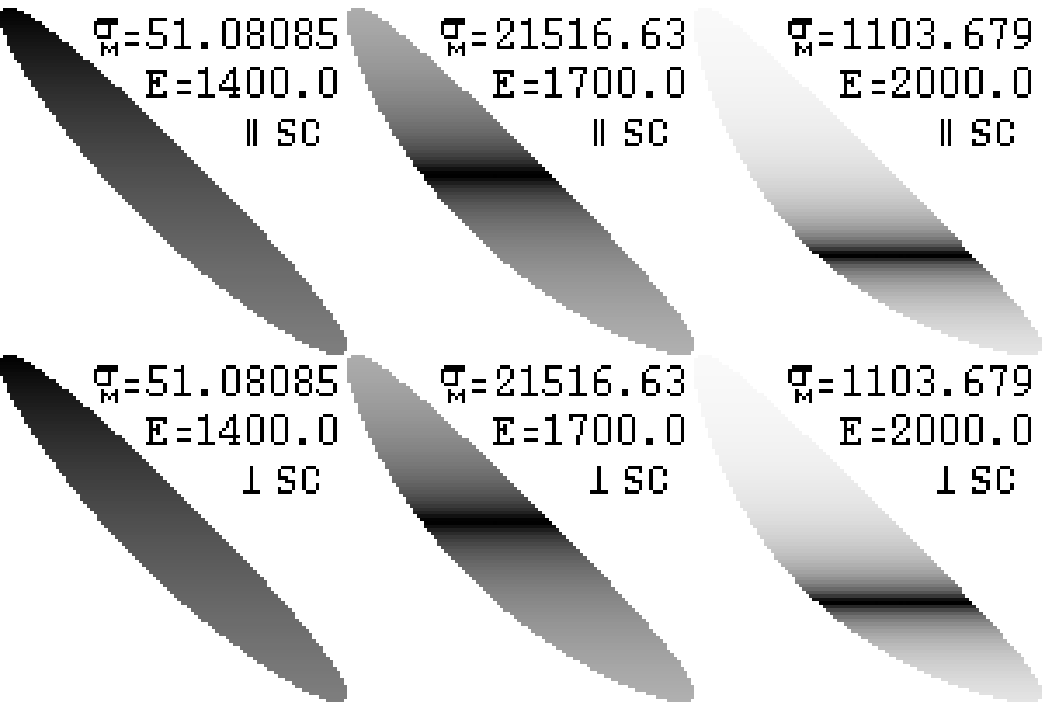}}
\caption{The Dalitz plots for the $S$ and $SC$ processes, the top number is
the maximum value of the invariant amplitude squared, the next number
is the energy in the center-of-mass frame, $\sqrt{s}$, and $\parallel$
or $\perp$ indicate the direction of the incoming nucleon momenta with respect
to the outgoing nucleon momenta}\label{fig3}
\end{figure}

\begin{figure}
\centerline{
\includegraphics[width=8cm]{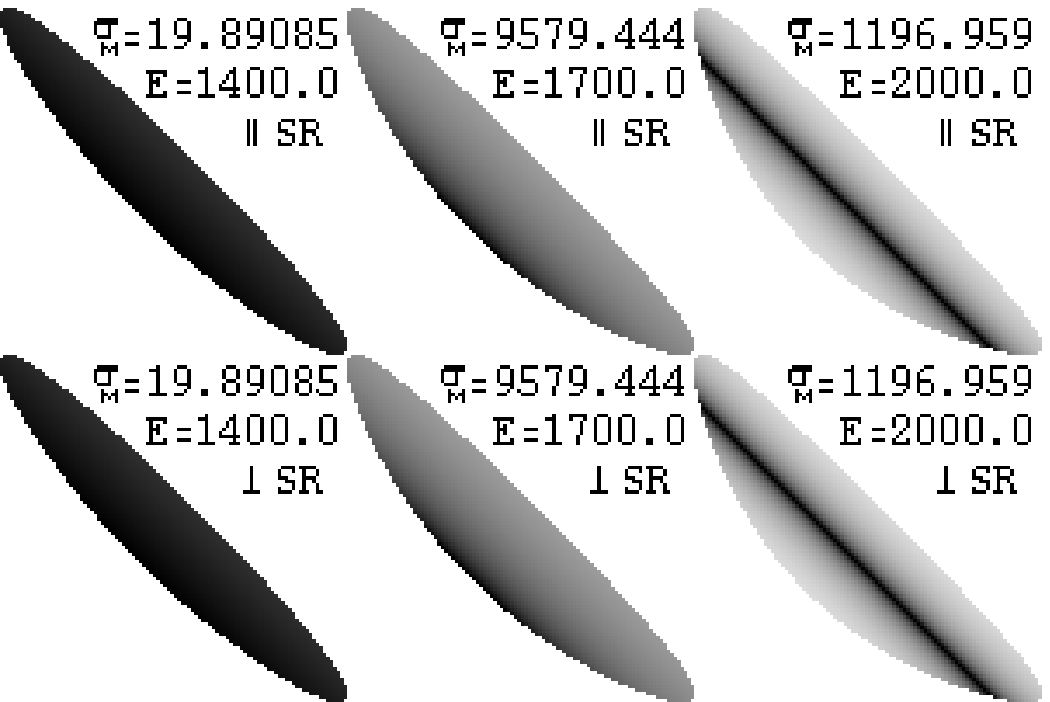}\includegraphics[width=8cm]{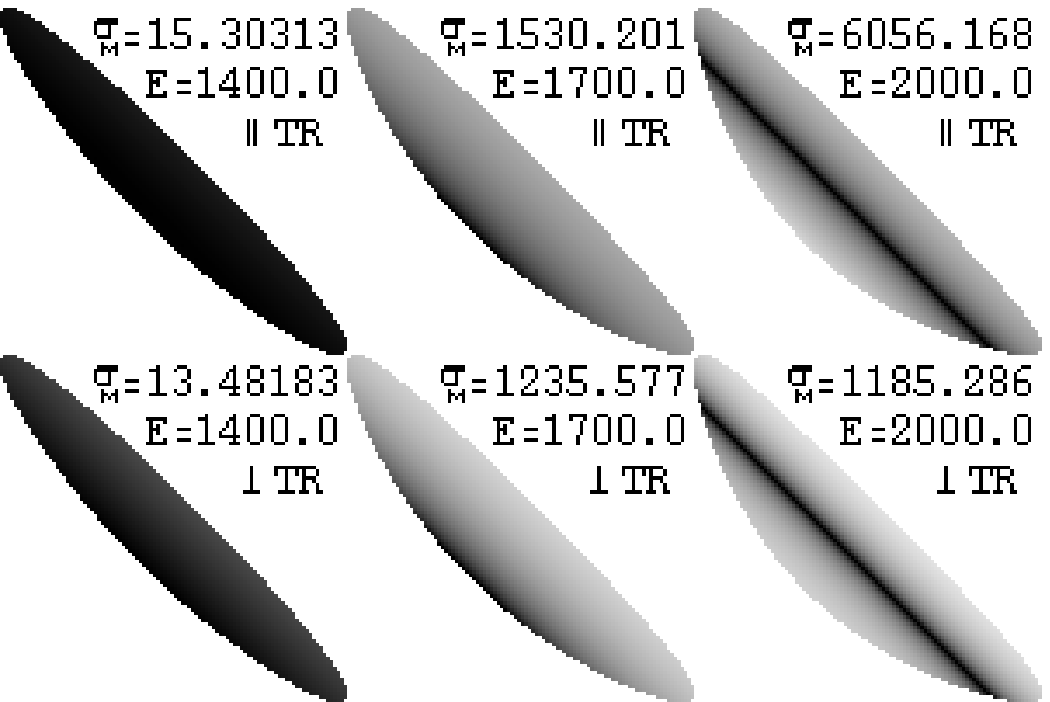}}
\caption{The Dalitz plots for the $SR$ and $TR$ processes.}\label{fig4}
\end{figure}

\begin{figure}
\centerline{
\includegraphics[width=8cm]{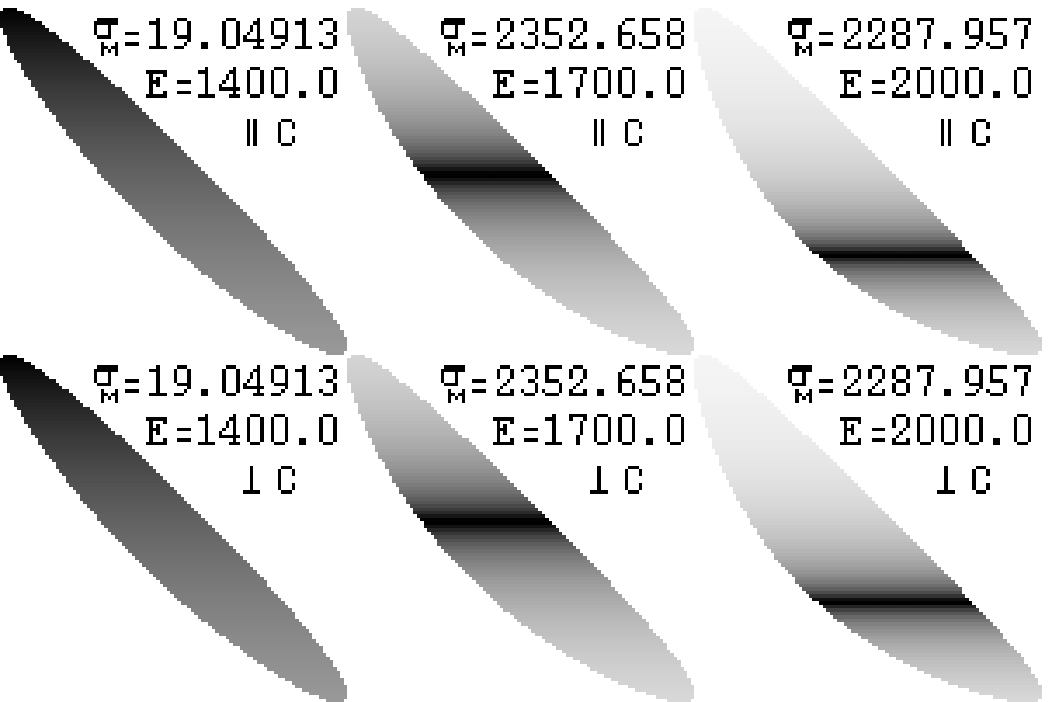}\includegraphics[width=8cm]{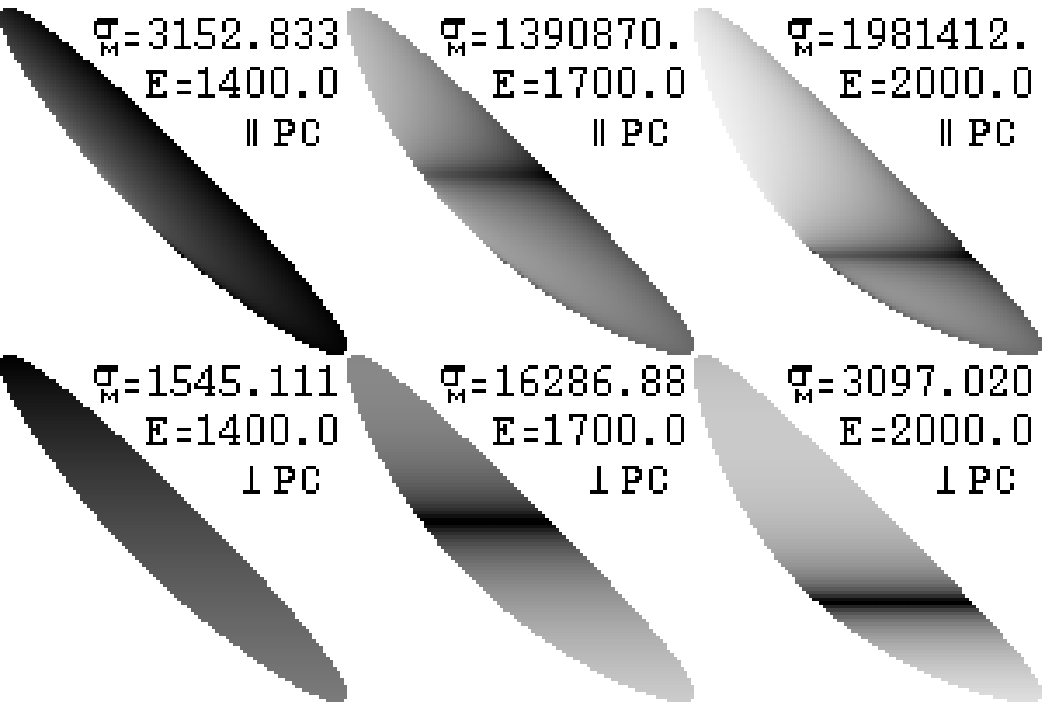}}
\caption{The Dalitz plots for the $C$ and $PC$ processes.}\label{fig5}
\end{figure}

\begin{figure}
\centerline{
\includegraphics[width=8cm]{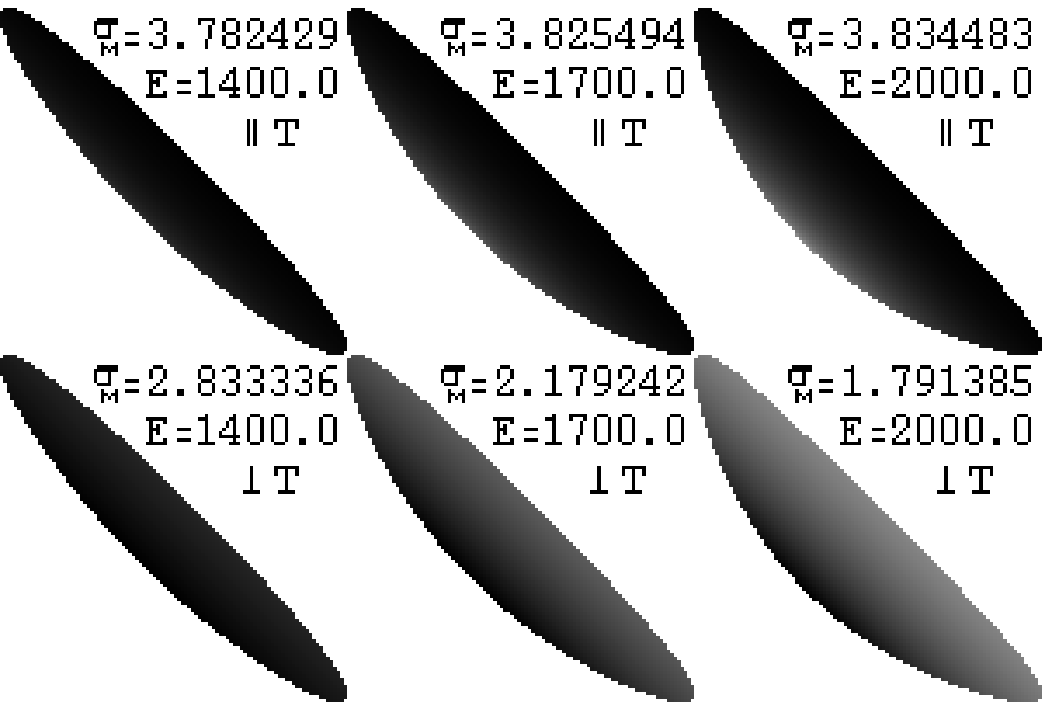}\includegraphics[width=8cm]{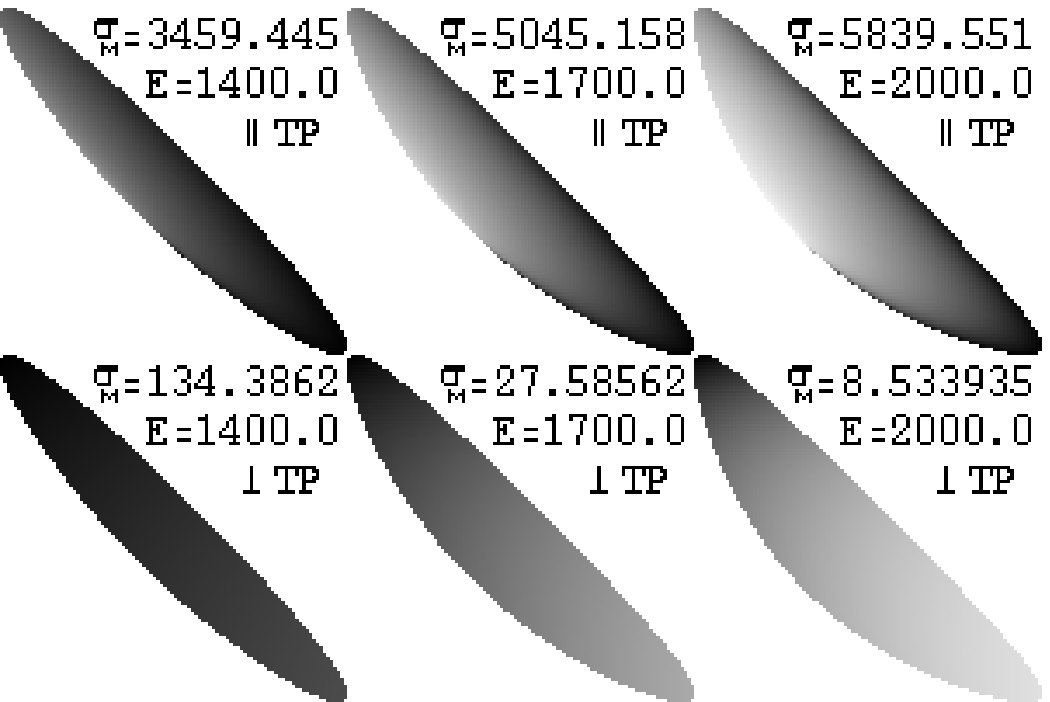}}
\caption{The Dalitz plots for the $T$ and $TP$ processes.}\label{fig6}
\end{figure}

\begin{figure}
\centerline{
\includegraphics[width=8cm]{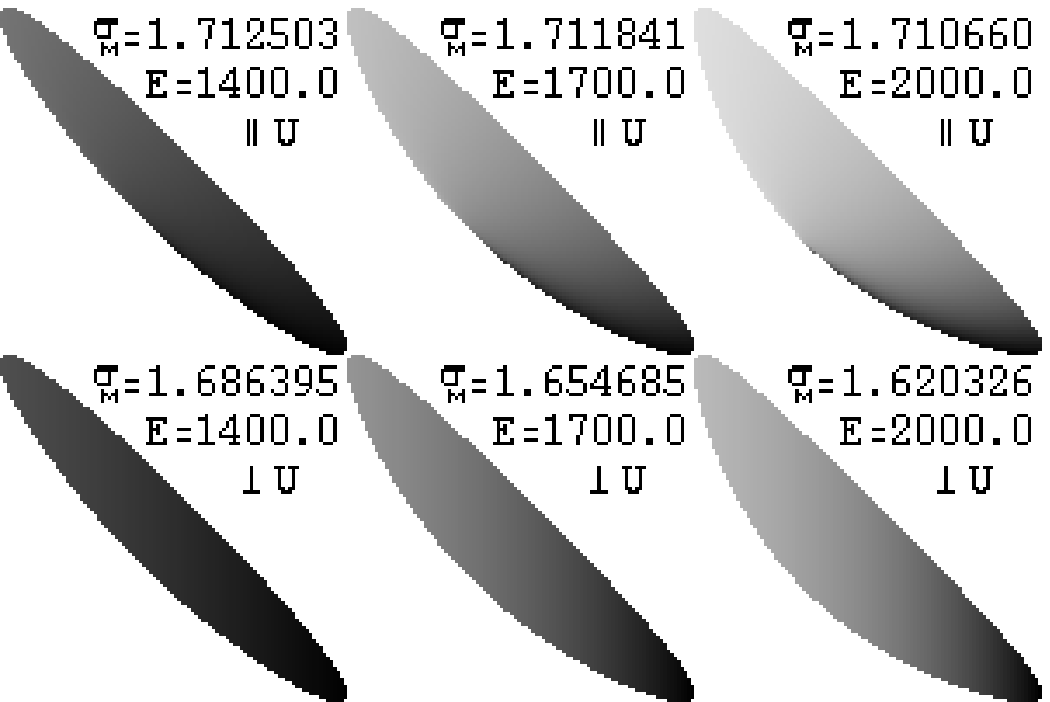}\includegraphics[width=8cm]{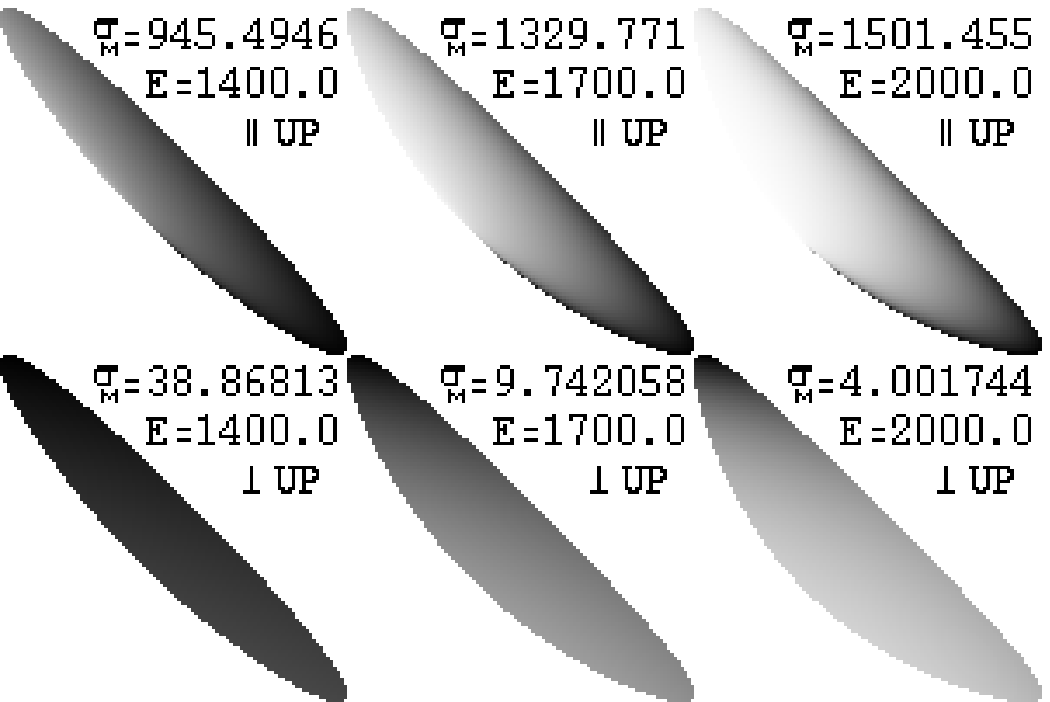}}
\caption{The Dalitz plots for the $U$ and $UP$ processes.}\label{fig7}
\end{figure}

\begin{figure}
\centerline{
\includegraphics[width=8cm]{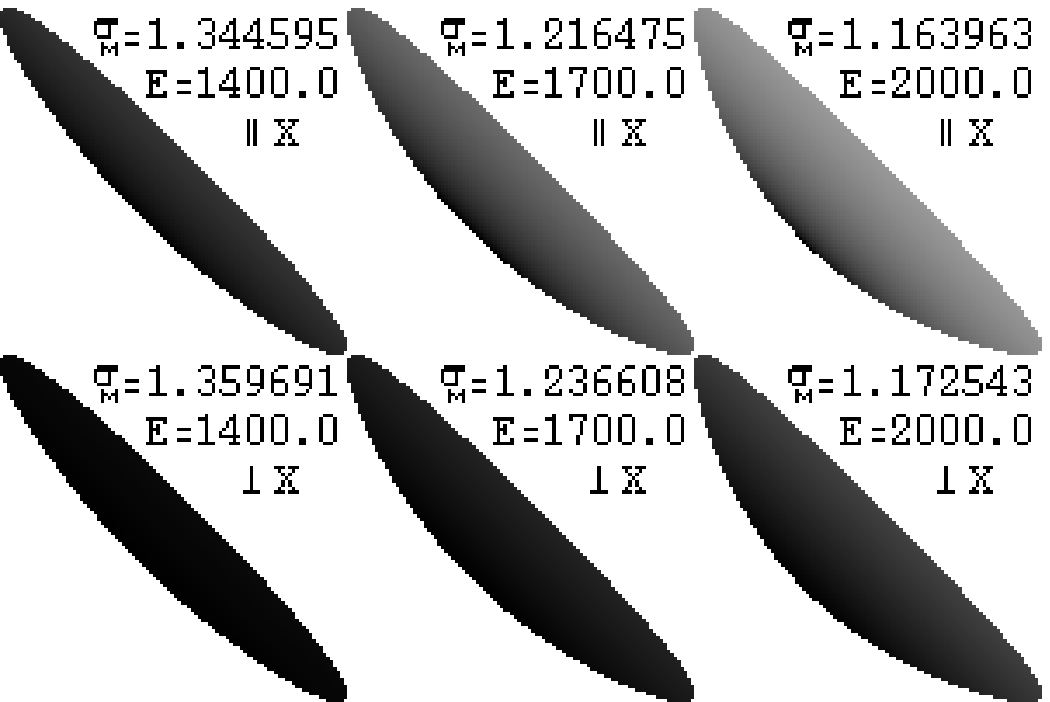}\includegraphics[width=8cm]{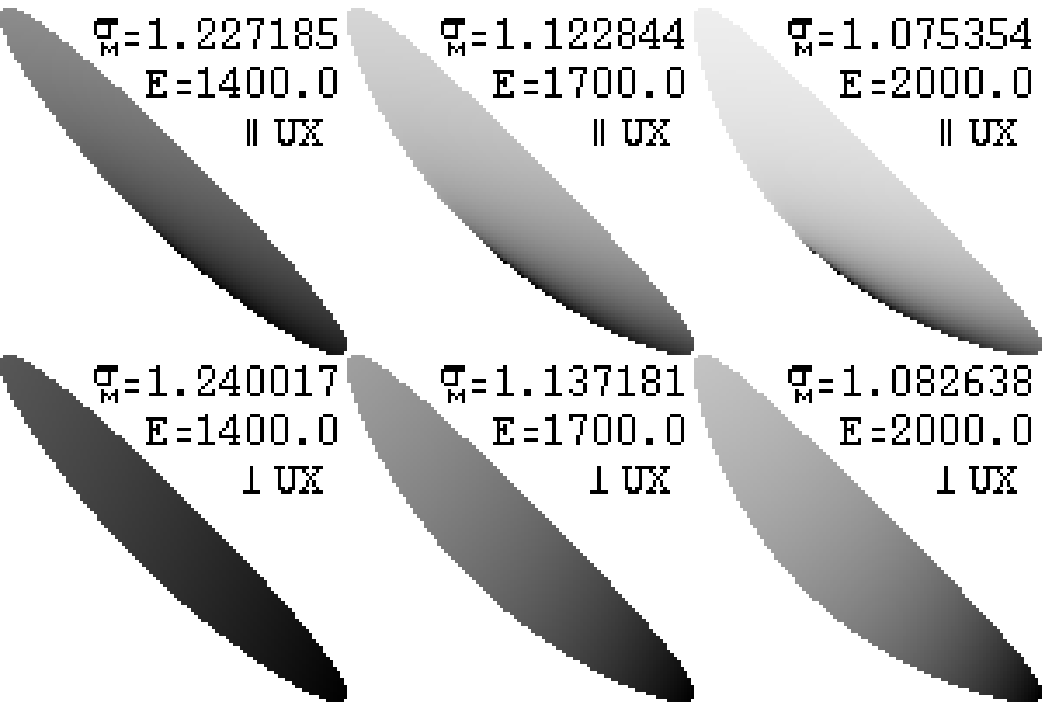}}
\caption{The Dalitz plots for the $X$ and $UX$ processes.}\label{fig8}
\end{figure}

\begin{figure}
\centerline{
\includegraphics[width=8cm]{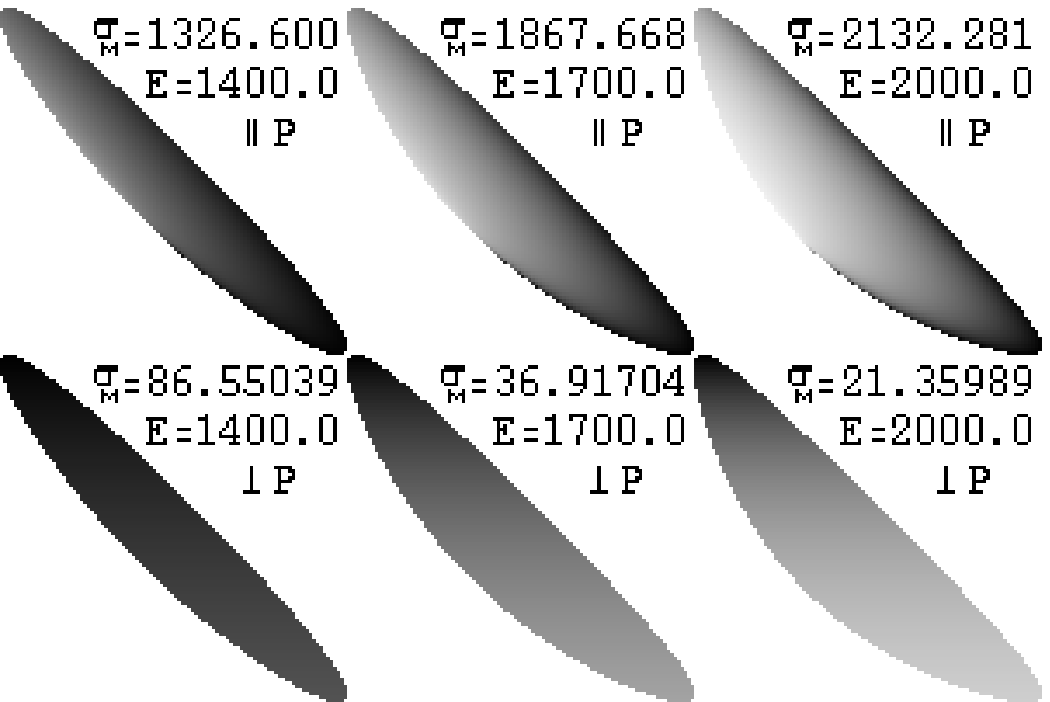}\includegraphics[width=8cm]{fig_C.eps}}
\caption{The Dalitz plots for the $P$ and $C$ processes.}\label{fig9}
\end{figure}

\begin{figure}
\centerline{
\includegraphics[width=8cm]{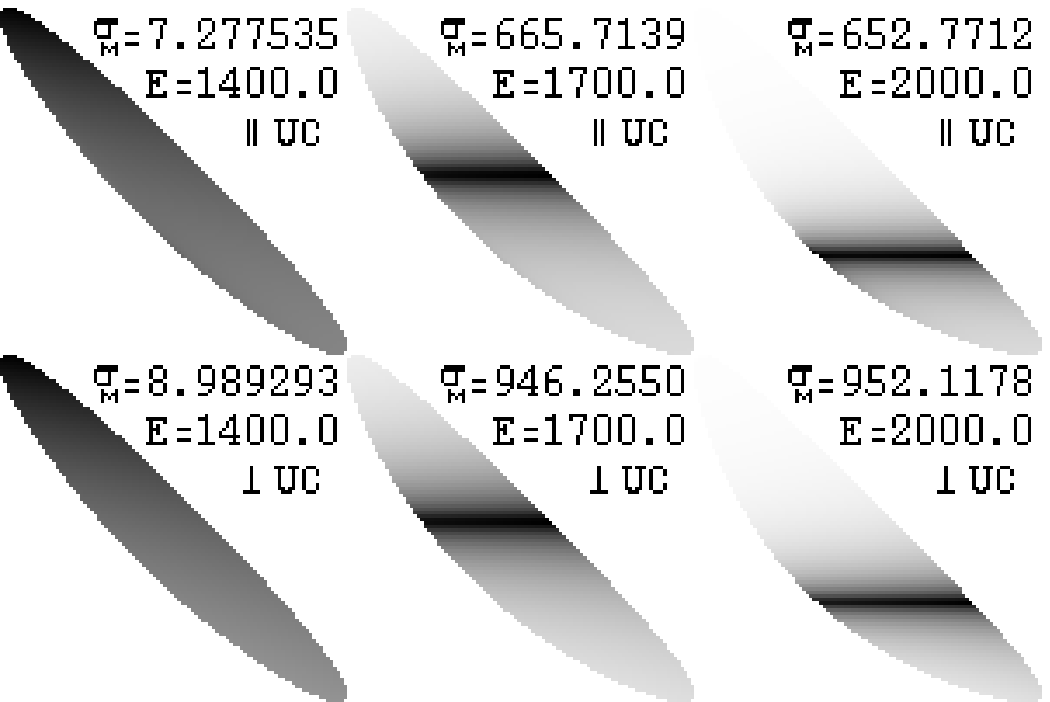}\includegraphics[width=8cm]{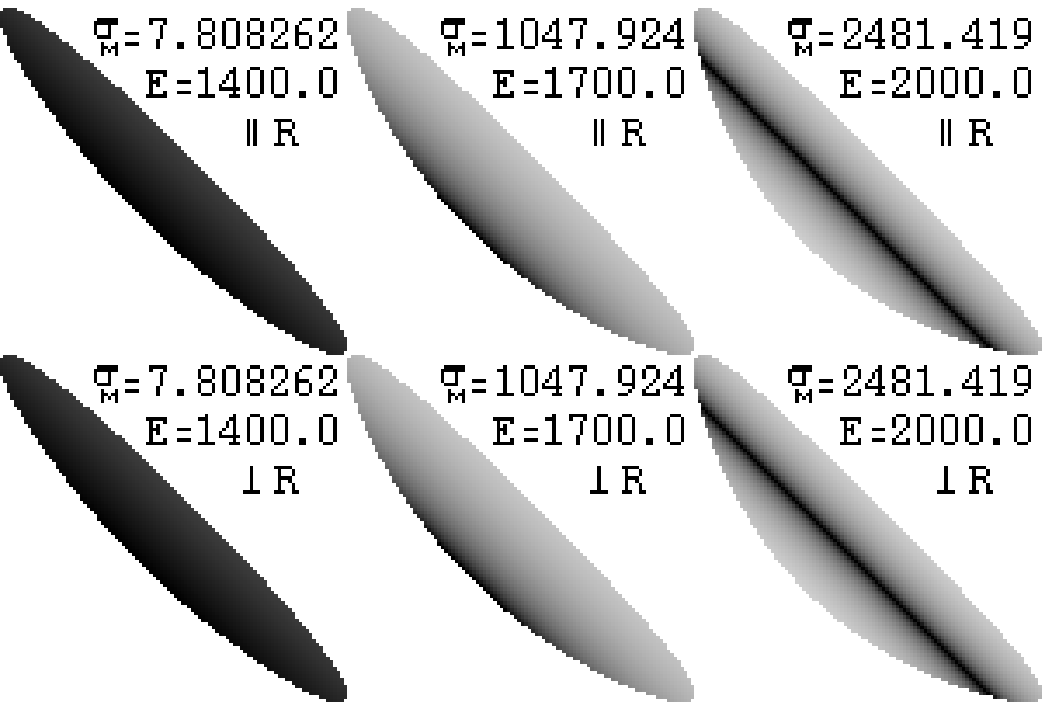}}
\caption{The Dalitz plots for the $UC$ and $R$ processes.}\label{fig10}
\end{figure}

\begin{figure}
\centerline{
\includegraphics[width=8cm]{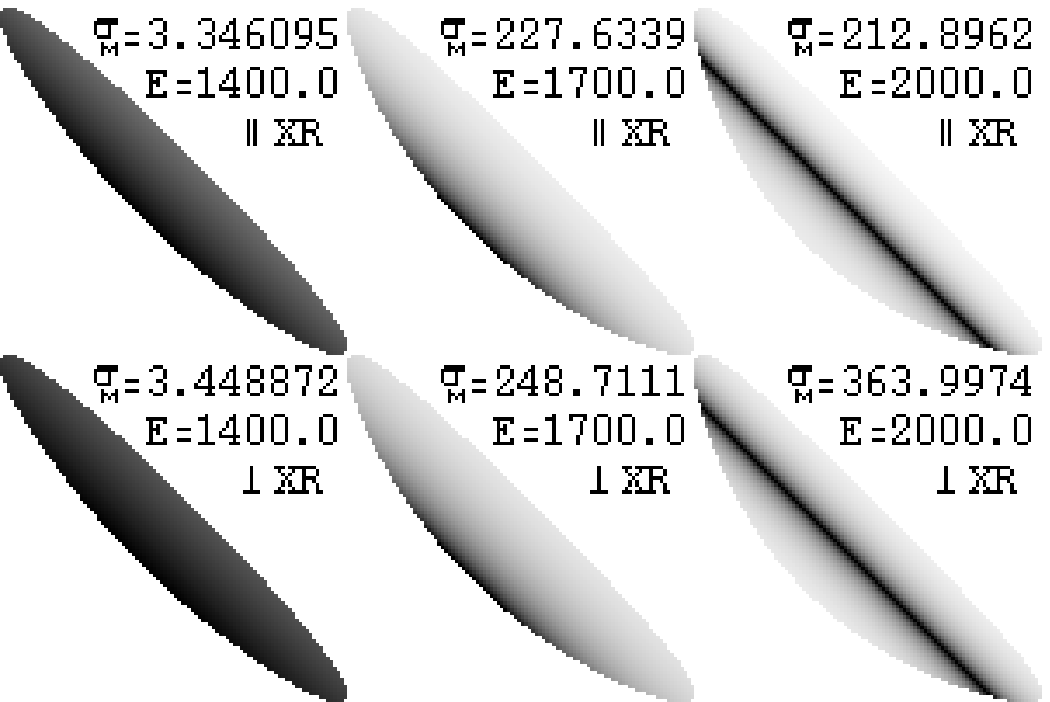}}
\caption{The Dalitz plots for the $XR$ process.}\label{fig11}
\end{figure}

We calculated the invariant amplitude squared $\sigma_M$
on a $100 \times 100$ grid
for each of the seventeen diagrams for the two angular cases 
and three values
of the invariant mass $E=\sqrt{s}$. The values we inserted
are not necessarily realistic, but archetypical:
$M_s = 1.6$GeV, $M_c = 1.35$GeV, $m_r = 0.77$GeV, 
and $m_\pi = 0.14$GeV. The widths are all set to
$\sqrt{\Gamma} = 0.14$GeV, except the pion which has
a small width of $\sqrt{\epsilon} = 0.003$GeV. 
It takes about five seconds to calculate these 102 Dalitz 
plots independently on a normal desktop computer, which makes a fitting
procedure in this approach feasible. The results are 
shown in Figs.~\ref{fig3}-\ref{fig11}, the trivial case
of ``1'' is not given.

The $S$ process gives a flat invariant amplitude across
the Dalitz plot. The $s$-resonance can be recovered
from comparing different scattering energies.
All the $S\ast$ processes have no angular dependence as
expected, so do the $R$ and $C$ processes. 
Vertex functions will introduce angular dependence
in the latter cases. The 
$U$ and $X$ processes have signatures that
depend on the angle, however, they are not particularly
peaked for parallel kinematics. Unlike the $P$
process, which is largely in the forward 
direction, where the intermediate pion momentum is
close to shell, which is also the reason for the large
values for this amplitude. The $T$ process is somewhere
in between: it has a larger amplitude in
parallel kinematics. Only the $S$, the $C$, and the $R$
processes give a clear resonance structure. It
is clear why many theoretical studies focus on these
diagrams, including combinations like $SC$, and $SR$. 
However, from this kinematical study
it is clear that none of the other processes is
obviously suppressed. One of the most complete studies
by Murphy and Laget~\cite{Murphy:1996ms} included
the diagrams: $C_\Delta,$
$C_{N^\ast},$
$(U_\Delta C_\Delta),$
$(S_p C_\Delta),$
$(S_{D_{13}} C_\Delta),$
$(S_{P_{11}} C_\Delta),$
$(S_{D_{13}} R_\rho),$
$(S_{P_{11}} R_\sigma),$
$(P C_\Delta),$
$(P C_{N^\ast}),$
$R_\rho,$
and $(T_{\rm Pomeron} R_\rho)$, which is clearly still a
small subset of all diagrams. In some two-body baryon 
resonance analysis effective two-pion states are included,
labeled by their angular momentum $\sigma (S=0)$ or $\rho(S=1)$.
It would be appropriate to check these channels against
the data. However, even from this small kinematical study
it is clear that this is not a trivial task.

Furthermore, assuming that $SC$ and $C$ and the dominant 
parts of the amplitude, we explored possible non-trivial
interference in the amplitude. However, in the case of 
just the single product of two production amplitudes
nothing unexpected happens; the interference is more
or less the product of the two amplitudes.

\section{Conclusions}

For the analysis of the data the complexity is
significantly reduced if the data is fitted only with
arbitrary functions $S,T,U,C,X,R,P$, possibly
separated in partial waves for $S$ and $C$.
However, it is clear that partial wave analysis
might have to yield to combined analysis in problems
like these.
To first order, the functions are sums of simple single
channel resonances, which masses and widths should 
be recovered, taking care of all the interference
among the different parts. The angular information
turned out to be important to separate the large
non-resonant background from the closed channels.

However, the widths are the direct result of the 
coupling of a particular resonance to its decay 
channels. In a proper analysis the coupling
strengths and the widths are highly correlated.
Therefore, in order to recover this consistency
one has to go beyond the perturbation theory
described in this paper and typically applied
to this problem, and include final-state
and intermediate-state interactions in
the analysis. This analysis is under investigation.
The results presented in this paper, of all
the possible production processes, are a first
step in this analysis.

\end{document}